\documentclass{article}
\usepackage[T1]{fontenc}
\usepackage{spconf,amsmath,graphicx,booktabs,tabularx,balance}
\usepackage[table]{xcolor}
\usepackage{hyperref}
\definecolor{resultshade}{gray}{0.94}
\title{From Reliable Text to Real Voices: Trust-Aware Progressive \\ Adaptation for Low-Resource TTS}
\name{\shortstack{Jiayi Lu$^{1,2,*}$, Yizhong Geng$^{1,3,*}$, Jinghan Yang$^{3}$,\\
Tianhan Jiang$^{4}$, Boxun An$^{5}$, Yingming Gao$^{3}$, Ya Li$^{3,\dagger}$}%
\thanks{$^{*}$Equal contribution. $^{\dagger}$Corresponding author: Ya Li.}}
\address{$^{1}$Beijing Logic Intelligence Technology \quad
$^{2}$University of Washington\\
$^{3}$Beijing University of Posts and Telecommunications\\
$^{4}$University of California, USA \quad $^{5}$Northwestern University, USA}
\begin{document}
\ninept
% Keep author notes and table notes at the template's 9 pt body size.
\let\footnotesize\small
\raggedbottom
\maketitle
\begin{abstract}
Low-resource text-to-speech (TTS) adaptation is constrained by scarce paired
data and costly manual transcription. Existing fixed-voice TTS systems can
provide relatively accurate pronunciation, but their synthetic speech offers
limited speaker diversity and may exhibit flat prosody. Real recordings
provide natural prosody and diverse voices, yet their automatic speech
recognition (ASR) pseudo-labels may contain transcription errors.
We find that supervision order affects content accuracy and speaker similarity.
We propose trust-aware progressive adaptation: synthetic-to-real adaptation
first establishes text--speech correspondences, then restores
reference-speaker control using real speech. Transcript-agreement weighting
uses agreement between two fixed ASR systems as a proxy for pseudo-label
reliability to limit noisy supervision. Experiments with FireRedTTS3 on
Burmese and Lao and OmniVoice on Burmese show improved content accuracy
with high naturalness and competitive speaker similarity. Jointly considering
supervision order and pseudo-label reliability when combining synthetic and
real speech offers a practical path to zero-shot voice cloning in low-resource
languages with less manual transcription.
Audio demos are available at
\url{https://insiderx-pro.github.io/S2R-Adaptation-TTS/}.
\end{abstract}

\section{Introduction}
\label{sec:intro}
Multilingual TTS enables zero-shot voice cloning from short reference
recordings~\cite{casanova2022yourtts,casanova2024xtts,le2023voicebox,zhang2023vallex}.
Recent systems broaden acoustic-domain and language
coverage~\cite{du2025cosyvoice3,zhu2026omnivoice,liu2025crosslingualf5}, but low-resource adaptation
still lacks high-quality text--speech pairs~\cite{lux2022lowresource,geng2025thai}.
Manual transcription is costly, motivating the use of unpaired text and
speech~\cite{chung2019semisupervised,makishima2022speaker}.

These resources offer two routes to constructing training pairs.
Existing TTS systems synthesize speech from text~\cite{joshi2023rapid},
providing relatively accurate pronunciation but potentially flat
prosody~\cite{geng2026stability}.
Fixed voices also limit speaker diversity, as illustrated by the
language-specific models in MMS~\cite{pratap2023mms}.
Synthetic speech thus provides useful pronunciation supervision but limited
acoustic diversity. Alternatively, ASR supplies pseudo-labels for real
speech~\cite{du2024cosyvoice2}, preserving natural prosody and diverse speaker
characteristics while introducing transcription noise.

These strengths motivate combining both sources, but synthetic
augmentation can degrade speaker similarity~\cite{choi2026zesta}.
Our controlled comparisons show that synthetic adaptation improves content
accuracy but can reduce speaker similarity; real speech restores
speaker similarity, but noisy pseudo-labels can erode content accuracy.
Reversing supervision order changes this trade-off, motivating joint
consideration of supervision order and pseudo-label reliability.

We propose trust-aware progressive adaptation.
Synthetic-to-real adaptation learns text--speech correspondences from
synthetic speech before using real speech to restore reference-speaker control.
Transcript-agreement weighting uses agreement between two fixed ASR systems
as a proxy for pseudo-label reliability.
We evaluate FireRedTTS3~\cite{shen2026fireredtts3} on Burmese and Lao and
OmniVoice on Burmese.

Our contributions are:
\renewcommand{\labelitemi}{$\bullet$}
\begin{itemize}
    \setlength{\itemsep}{2pt}
    \item \textbf{Analysis.} We characterize how supervision order affects
    content accuracy and speaker similarity.
    \item \textbf{Method.} We propose synthetic-to-real adaptation with
    transcript-agreement weighting for noisy pseudo-labels.
    \item \textbf{Validation.} We test effectiveness across languages and
    backbones using independent ASR and controlled interventions.
\end{itemize}

\section{Method}
\label{sec:method}
\begin{figure*}[t]
  \centering
  \includegraphics[width=\textwidth]{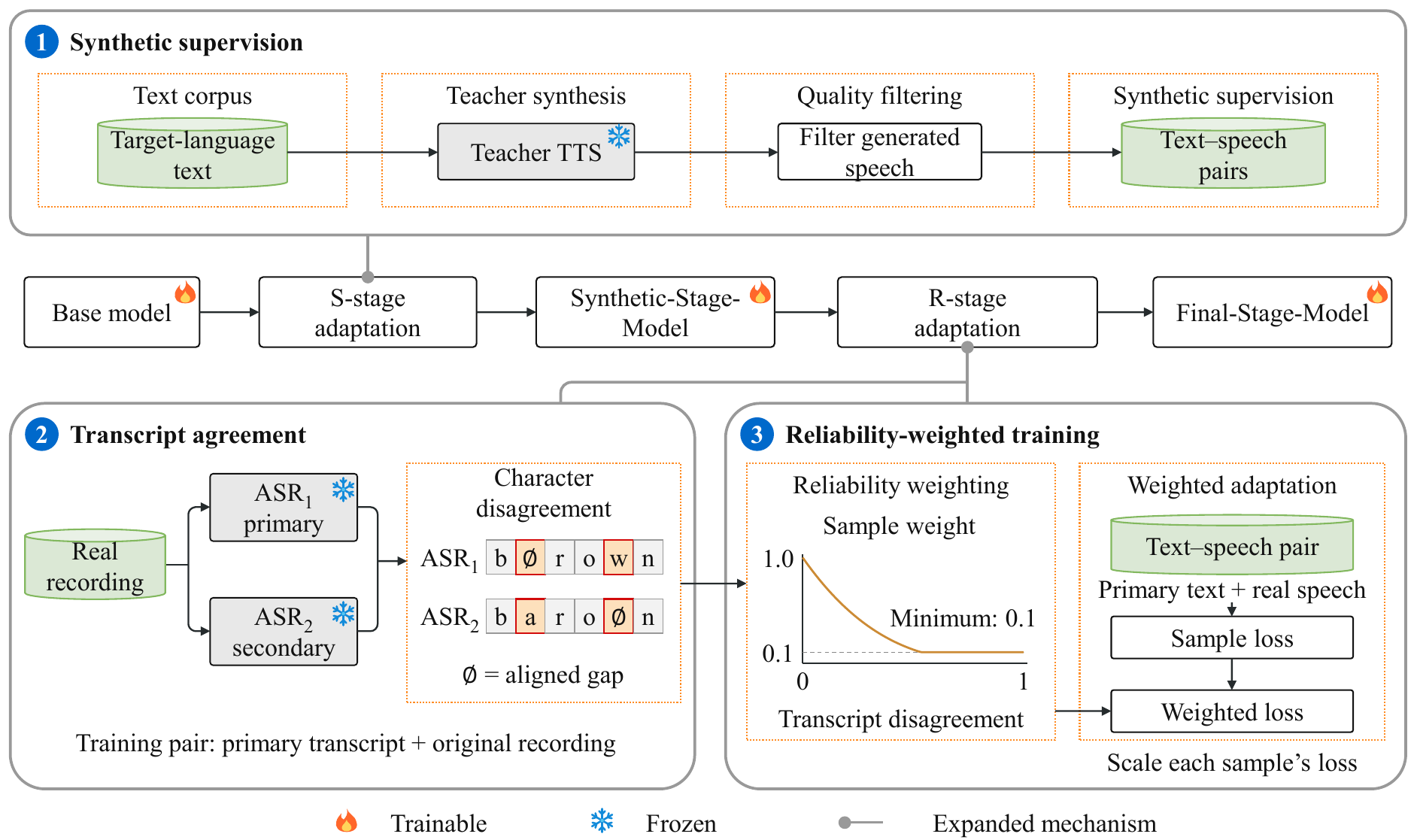}
  \caption{Trust-aware progressive adaptation. Synthetic supervision
  establishes text--speech correspondences in S-stage adaptation;
  transcript agreement guides reliability-weighted training on real
  recordings in R-stage adaptation. The primary transcript remains
  the training text.}
  \label{fig:pipeline}
\end{figure*}

In Fig.~\ref{fig:pipeline}, the Base model, Synthetic-Stage-Model, and
Final-Stage-Model are abbreviated Base, S, and S$\to$R:
synthetic supervision precedes reliability-weighted real supervision.

\subsection{Synthetic supervision}
Fixed Teacher TTS $T$ maps text $x_i$ and voice $v_i$ to
$y_i^S=T(x_i,v_i)$. Filtering (Section~\ref{sec:setup}) and same-voice
reference pairing yield $\mathcal D_S=\{(x_i,y_i^S,r_i^S)\}$, with
reference speech $r_i^S$, example index $i$, and S/R denoting
synthetic/real data. Training uses unit weights and original input text;
filtering does not guarantee correct pronunciation.

\subsection{Transcript agreement}
For real recording $y_i^R$, fixed primary ASR$_1$, $A_1$, supplies
$\tilde x_i=A_1(y_i^R)$; fixed secondary ASR$_2$, $A_2$, supplies only
a comparison transcript. Training uses
$\mathcal D_R=\{(\tilde x_i,y_i^R,r_i^R)\}$, without transcript fusion
or correction. Reference speech $r_i^R$ is a distinct 1--10 s utterance
from the same local speaker within a recording.

We measure normalized character disagreement:
\begin{equation}
d_i=\frac{\operatorname{ED}(\mathcal N(\tilde x_i),\mathcal N(A_2(y_i^R)))}
 {\max(1,|\mathcal N(\tilde x_i)|)}.
\label{eq:disagreement}
\end{equation}
$\operatorname{ED}$ and $|\cdot|$ denote code-point Levenshtein distance
and text length.
\par
\noindent
Scoring-only normalization $\mathcal N$ applies Unicode NFC,
removing whitespace and punctuation/symbol/control categories but retaining
letters, digits, and combining marks, without case, numeral, or Zawgyi conversion.
Empty primary labels are excluded; valid empty secondary transcripts give
$d_i=1$, distinct from failed ASR requests.

Agreement does not verify correctness: recognizers may share errors.
In Fig.~\ref{fig:pipeline}, character/transcript disagreement denotes $d_i$;
$\emptyset$ marks an alignment gap, not a training character.

\begin{table*}[t]
\centering
% Table~\ref{tab:main}. Use inside a table* in the original conference source.
% All numerical entries are from Template.pdf (the six-page revision).
% Display precision: CER 2 decimals; SIM-O 4 decimals; H and MOS unchanged.
\begingroup
\setlength{\aboverulesep}{1.2pt}
\setlength{\belowrulesep}{1.6pt}
\setlength{\tabcolsep}{1.5pt}
\renewcommand{\arraystretch}{1.12}
\vspace{-9pt}
\caption{Main results. (a) CER (\%), SIM-O, and $H$ (0--100); adapted CER/SIM-O: mean $\pm$ sample SD over three seeds. (b) MOS (1--5): mean [approximate 95\% crossed-bootstrap CI]. (c) Independent-ASR CER (\%). Bold: numerical bests, not significance.}
\label{tab:main}
\begin{tabular*}{\textwidth}{@{\extracolsep{\fill}}l*{3}{rrr}@{}}
\toprule
& \multicolumn{3}{c}{FireRedTTS3 / Burmese} & \multicolumn{3}{c}{FireRedTTS3 / Lao} & \multicolumn{3}{c}{OmniVoice / Burmese} \\
\cmidrule(lr){2-4}\cmidrule(lr){5-7}\cmidrule(l){8-10}
(a) Strategy & CER $\downarrow$ & SIM-O $\uparrow$ & $H\uparrow$ & CER $\downarrow$ & SIM-O $\uparrow$ & $H\uparrow$ & CER $\downarrow$ & SIM-O $\uparrow$ & $H\uparrow$ \\
\midrule
Base & 142.52 & 0.7244 & 0.00 & 100.45 & 0.7343 & 0.00 & 10.35 & 0.7278 & 80.34 \\
S & \mbox{17.43\,$\pm$\,0.45} & \mbox{0.4513\,$\pm$\,0.0041} & 58.36 & \mbox{13.49\,$\pm$\,0.26} & \mbox{0.6108\,$\pm$\,0.0006} & 71.61 & \mbox{8.21\,$\pm$\,0.20} & \mbox{0.6864\,$\pm$\,0.0010} & 78.54 \\
R & \mbox{54.09\,$\pm$\,0.68} & \mbox{\textbf{0.7314}\,$\pm$\,0.0005} & 56.41 & \mbox{37.06\,$\pm$\,0.36} & \mbox{\textbf{0.7537}\,$\pm$\,0.0019} & 68.60 & \mbox{10.74\,$\pm$\,0.41} & \mbox{\textbf{0.7328}\,$\pm$\,0.0013} & 80.48 \\
R$\rightarrow$S & \mbox{\textbf{14.71}\,$\pm$\,0.62} & \mbox{0.3636\,$\pm$\,0.0038} & 50.98 & \mbox{\textbf{10.64}\,$\pm$\,0.09} & \mbox{0.5959\,$\pm$\,0.0014} & 71.50 & \mbox{7.36\,$\pm$\,0.12} & \mbox{0.6948\,$\pm$\,0.0035} & 79.41 \\
S$\rightarrow$R: 1.0 & \mbox{23.27\,$\pm$\,0.15} & \mbox{0.6942\,$\pm$\,0.0014} & 72.89 & \mbox{14.99\,$\pm$\,0.25} & \mbox{0.6993\,$\pm$\,0.0010} & 76.74 & \mbox{10.21\,$\pm$\,0.22} & \mbox{0.7142\,$\pm$\,0.0017} & 79.56 \\
S$\rightarrow$R: 0.5 & \mbox{20.91\,$\pm$\,0.37} & \mbox{0.6996\,$\pm$\,0.0005} & 74.24 & \mbox{14.60\,$\pm$\,0.14} & \mbox{0.6968\,$\pm$\,0.0026} & 76.74 & \mbox{8.03\,$\pm$\,0.33} & \mbox{0.7094\,$\pm$\,0.0009} & 80.10 \\
\midrule
\textbf{S$\rightarrow$R: cubic} & \mbox{16.90\,$\pm$\,0.26} & \mbox{0.6997\,$\pm$\,0.0016} & \textbf{75.97} & \mbox{13.97\,$\pm$\,0.36} & \mbox{0.6968\,$\pm$\,0.0045} & \textbf{77.00} & \mbox{\textbf{6.85}\,$\pm$\,0.08} & \mbox{0.7098\,$\pm$\,0.0026} & \textbf{80.57} \\
\bottomrule
\end{tabular*}
\par\vspace{3pt}
\begin{tabular*}{\textwidth}{@{\extracolsep{\fill}}l*{3}{rr}@{}}
\toprule
& \multicolumn{2}{c}{FireRedTTS3 / Burmese} & \multicolumn{2}{c}{FireRedTTS3 / Lao} & \multicolumn{2}{c}{OmniVoice / Burmese} \\
\cmidrule(lr){2-3}\cmidrule(lr){4-5}\cmidrule(l){6-7}
Strategy & (b) MOS $\uparrow$ & (c) CER $\downarrow$ & (b) MOS $\uparrow$ & (c) CER $\downarrow$ & (b) MOS $\uparrow$ & (c) CER $\downarrow$ \\
\midrule
Base & 2.17\;[1.97,\,2.38] & 103.56 & 2.55\;[2.34,\,2.76] & 100.45 & 4.40\;[4.23,\,4.56] & 12.62 \\
S & 3.15\;[2.97,\,3.34] & 20.45 & 3.39\;[3.22,\,3.56] & 18.03 & 4.23\;[4.05,\,4.40] & 11.63 \\
R & 1.87\;[1.69,\,2.06] & 68.64 & 2.12\;[1.93,\,2.32] & 44.94 & 2.35\;[2.15,\,2.56] & 12.81 \\
R$\rightarrow$S & 2.72\;[2.52,\,2.92] & \textbf{17.93} & 3.27\;[3.07,\,3.47] & \textbf{17.13} & 4.32\;[4.12,\,4.50] & 11.65 \\
S$\rightarrow$R: 1.0 & 3.78\;[3.55,\,4.00] & 26.82 & 3.78\;[3.61,\,3.94] & 23.85 & 4.35\;[4.17,\,4.51] & 11.69 \\
S$\rightarrow$R: 0.5 & 3.91\;[3.71,\,4.11] & 25.06 & 3.84\;[3.65,\,4.02] & 23.33 & 4.42\;[4.24,\,4.58] & 11.41 \\
\midrule
\textbf{S$\rightarrow$R: cubic} & \textbf{4.12}\;[3.93,\,4.30] & 19.34 & \textbf{3.97}\;[3.77,\,4.17] & 18.53 & \textbf{4.51}\;[4.35,\,4.66] & \textbf{9.83} \\
\bottomrule
\end{tabular*}
\endgroup
\end{table*}

\subsection{Reliability-weighted training}
\textbf{Reliability weighting.} To limit noisy supervision's
contribution~\cite{ren2018reweight}, we assign
% Allow this displayed equation to continue in the next column.
\begingroup
\predisplaypenalty=0
\begin{equation}
w_i^{(\gamma)}=\max\!\left(w_{\min},[1-\min(d_i,1)]^\gamma\right),
\label{eq:weight}
\end{equation}
\endgroup
where $\gamma>0$ controls decay and $w_{\min}$ floors weights.
Cubic uses $\gamma=3$, $w_{\min}=0.10$ (Section~\ref{sec:setup}):
perfect agreement gets unit weight; disagreement reduces influence.

\textbf{Weighted adaptation.} For minibatch $\mathcal B$ and sample loss
$\ell_\theta$ at trainable parameters $\theta$,
\begin{equation}
\mathcal L_R(\theta)=\frac{1}{|\mathcal B|}
\sum_{i\in\mathcal B}w_i^{(\gamma)}
\ell_\theta(y_i^R,\tilde x_i,r_i^R),
\label{eq:loss}
\end{equation}
averages by batch size $|\mathcal B|$, not weight sum. Weighting changes
loss scale but need not proportionally scale AdamW updates. Offline
scoring adds no inference-time ASR.
Let $\mathcal A_B(\theta;\mathcal D,w)$ perform $B$ updates from $\theta$
on data $\mathcal D$ weighted by $w$. From Base $\theta_0$,
$\theta_S=\mathcal A_{B_S}(\theta_0;\mathcal D_S,\mathbf 1)$ and
$\theta_{SR}^{(\gamma)}=\mathcal A_{B_R}(\theta_S;\mathcal D_R,w^{(\gamma)})$,
where $B_S/B_R$ are update budgets and $\mathbf 1$/$w^{(\gamma)}$ collect
unit/real-example weights. Backbones/objectives stay unchanged;
stages restart optimizers/schedules. R starts from Base; R$\to$S reverses stages.

\subsection{Backbone objectives and reference conditioning}
FireRedTTS3~\cite{shen2026fireredtts3} trains its core with
RedAE/CAM++~\cite{wang2023campp} frozen. We weight its full
flow-matching~\cite{lipman2023flow}/stop loss,
$\ell_i=\ell_{\mathrm{flow},i}+0.1\ell_{\mathrm{stop},i}$, or OmniVoice's
masked acoustic-token loss~\cite{zhu2026omnivoice}.

RedAE/CAM++ (omitted from Fig.~\ref{fig:pipeline}) encode reference $r$
as acoustic prompt $z^r=E_{\mathrm{AE}}(r)$ and speaker embedding
$e^r=E_{\mathrm{spk}}(r)$. Controls fix pairings/reference text, exclude
ineligible targets, and pair mixture references within domains.
For text $x$ and $K$ patches~\cite{jia2025ditar,geng2026clasvs},
\begin{equation}
p_\theta(z_{1:K}\mid x,z^r,e^r)=
\prod_{k=1}^K p_\theta(z_k\mid x,z^r,e^r,z_{<k}),
\label{eq:history}
\end{equation}
History $z_{<k}$ is ground truth in training and generated at inference;
frozen encoders permit core-conditioning changes. For 16 texts, independent
prompt/embedding swaps use two same-text references. For generated $g$, define
$q=\mathrm{SIM}(g,r_A)-\mathrm{SIM}(g,r_B)$ (SIM-O cosines);
$\Delta_{\mathrm{emb}}$ averages the $r_B$-to-$r_A$ embedding-induced
change in $q$ over prompts/texts. History tests use 8/23 prefix pairs
(normalized ASR CER $\leq5\%$), fixing reference, boundary, and subsequent
random state. Suffix scoring excludes prefixes and the first 250 ms;
unchanged-history replay reproduces outputs (Section~\ref{sec:analysis}).

\newpage
\balance
\section{Experiments}\label{sec:setup}
\subsection{Experimental setup}
\textbf{Data.} Burmese S/R comprise 160,918 Azure Nilar/Thiha pairs (291.148 h) and 48,043 DVB pairs (92.263 h); Lao has 119,246 S and 26,004 R pairs (200.004/45.117 h). ASR$_1$ is Gemini 2.5 Flash; ASR$_2$ is Omnilingual ASR~\cite{omnilingual2025asr}. Synthetic filtering checks valid audio/text, 0.4--20 s duration, ASR CER $\leq20\%$, speech fraction $\geq0.65$, clipping $\leq0.005$, deduplication, evaluation exclusions, reference eligibility, and a 1,024-token limit. Burmese S starts from 171,240 candidates; 717 real targets lack references. Real speech spans 300 recordings, 1,200 local speaker groups, and 286 unverified cross-recording identity clusters.

\vspace{3.85pt}
\textbf{Training.} Burmese S/R use 53,640/16,015 updates at learning rates $2\times10^{-6}/10^{-6}$; Lao uses 39,749/8,668 updates. Orders share AdamW~\cite{loshchilov2019adamw}, 3\% warmup and cosine decay. All adaptation/control runs use one device, accumulation/effective batch 3, and training seeds 42/17/73. Tables~\ref{tab:main}a and~\ref{tab:weighting} report adapted CER/SIM-O as mean $\pm$ sample SD over these seeds; Base has no training-seed SD. Dev120 (60 texts $\times$ two references) selects $\gamma=3$, $w_{\min}=0.10$ by minimum three-seed mean CER with SIM-O $\geq0.6970$, over $\gamma\in\{0.5,1,2,3,4,6\}$, $w_{\min}\in\{0,0.05,0.10,0.20\}$. Broadcast R uses 16,015 updates, not test-selected.

\textbf{Evaluation.} Burmese Common400 (100 general/300 cloning conditions; 150 texts) uses Omnilingual ASR Unlimited 7B v2. Clone300 uses 30 Burmese/English/Chinese references; SIM-O averages ECAPA-TDNN/WavLM-large cosines~\cite{desplanques2020ecapa,chen2022wavlm}. Lao uses Gemini on 404 FLEURS~\cite{conneau2022fleurs} rows (260 inputs, one reference). CER comparisons are within-language. Common400 reuses FLEURS train/development conditions. Normalized evaluation texts do not overlap archived DVB or Lao S/R transcripts. Burmese hashes match none of the 100 checked target/reference conditions. Audited Clone300 speakers are disjoint from adaptation speakers; audio near-duplicate auditing remains incomplete. FireRedTTS3 decoding uses seed 42, 10 flow steps, guidance 2.0, and stop threshold 0.5; long/empty outputs remain scored. Joint score $H=200as/(a+s)$ uses $a=\operatorname{clip}(1-\mathrm{CER}/100,0,1)$, $s=\operatorname{clip}(\mathrm{SIM\mbox{-}O},0,1)$, and $H=0$ if $a+s=0$.

\textbf{Listening test.} Per language, 20 listeners rated all 30 matched text--reference conditions per system (naturalness 1--5; 600 ratings/system). Burmese backbones share listeners and conditions. Approximate 95\% percentile CIs use 10,000 crossed-bootstrap replicates~\cite{owen2007pigeonhole}, independently resampling listeners and conditions with common indices across systems within each language. We report paired cubic-minus-uniform-0.5 differences. Listening-test audio uses each adapted system's seed-42 checkpoint.

\begin{figure*}[t]
  \centering
  \includegraphics[width=\textwidth]{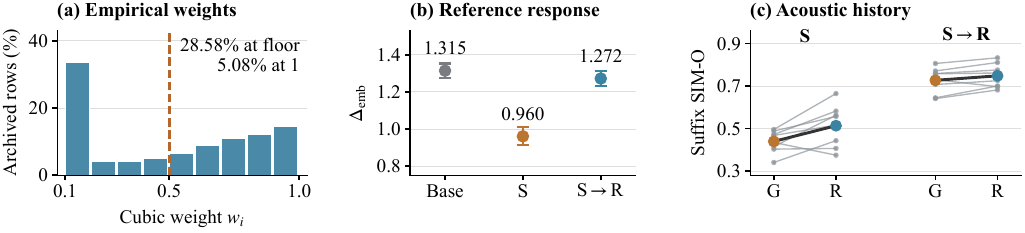}\par
  \vspace{-10pt}
  \caption{(a) Cubic weights: 42,785 archived ASR rows.
  (b) Embedding response: 16 texts, paired-bootstrap 95\% intervals.
  (c) Eight suffix SIM-O pairs: generated (G)/recorded (R) prefixes; thick lines: means.
  (b,c): historical uniform 0.5 S$\to$R, not cubic.}
  \label{fig:analysis}
\end{figure*}

\begin{table*}[t]
\nocite{dolphinsmall,sianglaoxlsr}
\noindent\begin{minipage}[t]{\columnwidth}
\begingroup
\setlength{\aboverulesep}{1.2pt}
\setlength{\belowrulesep}{1.6pt}
\setlength{\tabcolsep}{3pt}
\renewcommand{\arraystretch}{1.3684}
\vspace{-7pt}
\caption{Burmese systems on Common400 / Clone300. $^*$: fixed voices; their SIM-O scores are descriptive.}
\label{tab:baselines}
\begin{tabular*}{\columnwidth}{@{\extracolsep{\fill}}lrr@{}}
\toprule
System & CER (\%) $\downarrow$ & SIM-O $\uparrow$ \\
\midrule
Fish Audio S2-Pro~\cite{fish2026s2} & 83.89 & 0.6376 \\
MMS-TTS-mya$^*$~\cite{pratap2023mms} & 20.49 & 0.1029 \\
F5-Myanmar-TTS v2~\cite{f5myanmar2026,chen2025f5tts} & 35.24 & 0.5863 \\
IMS-Toucan~\cite{toucan2024repo} & 95.40 & 0.3298 \\
mmSpeech Tacotron$^*$~\cite{mmspeech2019} & 78.70 & 0.0914 \\
OmniVoice Base~\cite{zhu2026omnivoice} & 10.35 & \textbf{0.7278} \\
\midrule
FireRedTTS3 cubic & 16.90 & 0.6997 \\
\textbf{OmniVoice cubic} & \textbf{6.85} & 0.7098 \\
\bottomrule
\end{tabular*}
\endgroup
\end{minipage}\hfill
\begin{minipage}[t]{\columnwidth}
\begingroup
\setlength{\aboverulesep}{1.2pt}
\setlength{\belowrulesep}{1.6pt}
\setlength{\tabcolsep}{1.5pt}
\renewcommand{\arraystretch}{1.00}
\vspace{-7pt}
\caption{Burmese FireRedTTS3 controls from S. Mean $\pm$ sample SD over three seeds; shuffles are averaged within seed.}
\label{tab:labels}\label{tab:weighting}
\begin{tabular*}{\columnwidth}{@{\extracolsep{\fill}}lrr@{}}
\toprule
Setting & CER (\%) $\downarrow$ & SIM-O $\uparrow$ \\
\midrule
\multicolumn{3}{@{}l}{\textit{(a) FLEURS label control; S: Table~\ref{tab:main}a}} \\
Pseudo-labels, $w=0.5$ & \mbox{21.17\,$\pm$\,0.38} & \mbox{0.6345\,$\pm$\,0.0024} \\
Pseudo-labels, $w=1.0$ & \mbox{23.67\,$\pm$\,0.19} & \mbox{\textbf{0.6487}\,$\pm$\,0.0009} \\
Official text, $w=1.0$ & \mbox{\textbf{15.45}\,$\pm$\,0.15} & \mbox{0.6342\,$\pm$\,0.0012} \\
\midrule
\multicolumn{3}{@{}l}{\textit{(b) Broadcast weighting; uniform 1.0: Table~\ref{tab:main}a}} \\
Uniform 0.5 & \mbox{20.91\,$\pm$\,0.37} & \mbox{0.6996\,$\pm$\,0.0005} \\
Linear ($\gamma=1$) & \mbox{23.25\,$\pm$\,0.49} & \mbox{0.6956\,$\pm$\,0.0030} \\
Global shuffle & \mbox{21.34\,$\pm$\,0.52} & \mbox{0.6984\,$\pm$\,0.0010} \\
Duration-strat. shuffle & \mbox{21.13\,$\pm$\,0.22} & \mbox{0.6988\,$\pm$\,0.0032} \\
Uniform, equal mean & \mbox{21.09\,$\pm$\,0.55} & \mbox{0.6993\,$\pm$\,0.0008} \\
\midrule
\textbf{Cubic ($\gamma=3$)} & \mbox{\textbf{16.90}\,$\pm$\,0.26} & \mbox{\textbf{0.6997}\,$\pm$\,0.0016} \\
\bottomrule
\end{tabular*}
\endgroup
\end{minipage}
\end{table*}

\subsection{Main results}\label{sec:results}
Table~\ref{tab:main} shows order dependence, also in matched-batch Lao: uniform S$\to$R\ restores similarity but can worsen CER; R$\to$S\ favors CER. Cubic has the highest observed $H$ and mean MOS. Against uniform 0.5, paired MOS differences are $+0.210\,[0.025,0.392]$, $+0.130\,[-0.047,0.315]$, and $+0.090\,[-0.052,0.230]$ for FireRedTTS3/Burmese, FireRedTTS3/Lao, and OmniVoice/Burmese. Only the first marginal 95\% interval excludes zero; these exploratory comparisons are unadjusted for multiplicity. We do not claim consistent MOS improvement across settings. OmniVoice exceeds R in $H$ by just 0.08/100.

\textbf{Independent ASR.} Dolphin-small~\cite{dolphinsmall} (Burmese) and XLS-R Lao~\cite{sianglaoxlsr} (Lao) rescore the same generated audio and target texts. Neither participated in pseudo-labeling, reliability estimation, filtering, or model selection. Failures, empty transcripts, and overlong outputs are not selectively excluded. Cubic improves on uniform 1.0 in all groups (Table~\ref{tab:main}c); R$\to$S\ still yields lower FireRedTTS3 CER, consistent with an accuracy--similarity trade-off.

Adapted OmniVoice leads the shared-text Burmese comparison (Table~\ref{tab:baselines}), reducing Base CER by 3.50 points; fixed-voice SIM-O is descriptive. All outputs are scored, including 210/400 mmSpeech Tacotron cases reaching its 12.54 s limit and IMS-Toucan outputs with Burmese phoneme warnings.

\newpage
\nobalance
\section{Ablation and Analysis}\label{sec:analysis}
\vspace{10.17pt}
\subsection{Supervision order and pseudo-label reliability}
\textbf{Label quality.} Table~\ref{tab:labels}a compares labels on identical FLEURS recordings: 3,058 (12.135 h), retaining 2,921 (11.578 h) after overlap/duration exclusions. Official text reduces CER by 8.22 points at $w=1$, retaining similarity gains; halving pseudo-label weight helps less. Label noise contributes alongside other domain effects.

\textbf{Reliability validation.} On these 2,921 Burmese FLEURS utterances, ASR$_1$ pseudo-label CER against official text, $e_i$, correlates with $d_i$: Spearman $\rho=0.560$, Pearson $r=0.546$. Mean label CER is 5.94\% at $d_i<0.1$ ($n=971$) versus 32.30\% at $d_i\geq0.6$ ($n=286$). Agreement is informative, not proof of correctness.

\textbf{Weight assignment.} Table~\ref{tab:weighting}b fixes initialization, targets, references, order, and updates. Global shuffles of cubic weights (101/202/303 crossed with three training seeds) break assignment; duration-decile shuffles control length correlations. Both preserve weights yet, like equal-mean uniform weighting, underperform cubic: assignment matters beyond loss scale. The 42,785-row archive (Fig.~\ref{fig:analysis}a) has mean/SD 0.50166/0.33337 and quartiles 0.10000/0.54771/0.81090. Its 588 later-excluded rows and missing later transcripts distinguish it from the 48,043-target pool (mean 0.500946). The pool's duration-weighted mean 0.58715 becomes $0.50110/{\approx}0.5865$ with global/duration-stratified shuffling.

\textbf{Mixtures.} A 0/10/25/50/75/90/100\% real-example sweep from S uses 16,015 updates with unit S/cubic R weights; synthetic replay restores less similarity than real-only continuation. Joint training from Base matches 160,920 S/48,045 R exposures and the learning-rate schedule (restart at 53,640), but yields 20.160\% CER/0.65120 SIM-O, supporting progressive adaptation under matched exposure.

\newpage
\subsection{Reference conditioning and acoustic history}
\vspace{-3pt}
\textbf{Reference response.} Base/S/historical uniform 0.5 S$\to$R\ yield $\Delta_{\mathrm{emb}}$ of 1.3147/\allowbreak0.9602/\allowbreak1.2723 (Fig.~\ref{fig:analysis}b); matched-reference SIM-O is 0.7804/\allowbreak0.5672/\allowbreak0.7625. Conflicts follow embeddings; R restores S-weakened response; cubic is untested.

\textbf{History dependence.} Recorded versus generated prefixes improve suffix SIM-O by 0.0739/0.0211 for S/uniform 0.5 S$\to$R\ (Fig.~\ref{fig:analysis}c), in 7/8 cases each. Paired 95\% intervals $[0.0225,0.1204]$ and $[0.0025,0.0374]$ condition on cases, not seeds. This supports history dependence, not naturally accumulating voice drift.

\vspace{-6pt}
\section{Conclusion}
\vspace{-3pt}
We presented trust-aware progressive adaptation for low-resource TTS.
It combines a synthetic-to-real adaptation schedule with
transcript-agreement weighting.
Across the evaluated languages and backbones, synthetic speech improves
content accuracy but can reduce speaker similarity, while subsequent adaptation
on real speech restores reference-speaker control at the risk of transcription
noise. Transcript-agreement weighting helps preserve content accuracy,
with improvements over uniform weighting persisting under independent ASR
evaluation. Ground-truth label analysis supports transcript agreement as an
informative, imperfect reliability proxy, while conditioning interventions
reveal changes in reference response and sensitivity to acoustic history.
Together, these findings suggest that weak supervision should be organized
around the capabilities it strengthens or preserves, rather than treated as
interchangeable additional data. By combining the pronunciation supervision
available from existing TTS systems with the acoustic diversity of real speech,
this approach offers a practical path toward zero-shot voice cloning in
low-resource languages with reduced dependence on manual transcription of
target-language speech.

% Technical content, including floats, must end within the first four pages.
\clearpage
\balance
\bibliographystyle{IEEEbib}
\bibliography{refs}
\end{document}